# Determination of carbohydrates in infant milk powders by ultra-performance liquid chromatography with evaporative light scattering detector and BEH HILIC column

Hyon-Hui Ri[a*], Yong-A Choe[a], Jong-Ho Yun[a], Kye-Ryong Sin[b]

(a: Institute of Analysis and b: Faculty of Chemistry, **Kim Il Sung** University,

Pyongyang, Democratic People's Republic of Korea)

* E-mail: ryongnam17@yahoo.com

**Abstract**

The possibility of carbohydrate separation in BEH HILIC (Ethylene Bridged Hybride, Hydrophilic Interaction Liquid Chromatography) column was studied by ultra-performance liquid chromatography (UPLC) with evaporative light scattering detector (ELSD) and mobile phase containing amine compounds as modifiers. The chromatography conditions and ELSD parameters were optimized to separate five typical carbohydrates and applied to analysis of four infant milk powders. The linear ranges of carbohydrate determination were 20-300mg/L for fructose and glucose, 20-250mg/L for sucrose and lactose, and 35-180mg/L for fructo-oligosaccharide. The LODs were 16.4mg/L for fructose and glucose, 17.3mg/L for sucrose, 20.0mg/L for lactose, and 46.7mg/L for fructo-oligosaccharide. Relative standard deviations (RSDs) ranged between 3.45-4.23%, 1.46-4.17%, 4.14-5.60%, 1.39-4.09%, and 2.49-3.61% for fructose, glucose, sucrose, lactose, and fructo-oilgosaccharide, respectively and recoveries ranged between 95.0 and 105.4%

**Keyword**; ultra-performance liquid chromatography, evaporative light scattering detector, HILIC column, carbohydrate, milk powder

## 1. Introduction

Carbohydrates are important in nutritional quality assessment of food. There are many methods to analyze carbohydrates such as gas chromatography (GC) [1-3], ion chromatography (IC) [9,10], and capillary electrophoresis (CE) [11,12], but high-performance liquid chromatography (HPLC) has been mostly used [4-8]. For the analysis of carbohydrates by HPLC, refractive index detector (RID) and evaporative light scattering detector (ELSD) have been used as detector [5, 13, 14]. HPLC with RID has some disadvantages such as limited sensitivity and incompatibility with gradient eluent, compared with ELSD, which can be used to detect most of compounds with less volatile than mobile phase. The detection by ELSD is based on the versatility of particles to cause photon scattering. The ELSD is gradient compatible and has excellent baseline stability without the effects of mobile phase and temperatures. It has been widely used in determination of carbohydrates in different samples such as drinks, medicine, milk product, plants and so on [5].

In determination of carbohydrates by HPLC, an amine-bonded silica gel column was mostly used. But the separation of carbohydrates on the amine-bonded silica gel column is not always quantitative. This is due to the possible interaction between reducing carbohydrates and the amino group of the ligand, that is, due to formation of a Schiff's base and self-hydrolysis of the basic material [15]. Therefore, various HPLC techniques have been developed, which use silica columns [16, 17] and octadecylsilica columns [18] to separate the carbohydrates by adding amine compounds such as ethylendiamine [15], triethylamine [19], n-alkylamines to the mobile phase. The dynamically modified amino columns are more stable than the chemically-bonded amino columns in the separation, since they are continuously generated [15].

BEH HILIC column is excellent for retention of very polar, basic, water soluble analytes, since it is specifically designed and tested for HILIC separations using mobile phases containing high concentrations of organic solvent. Because it is similar to silica gel column in its structure and has silanol groups, it can be used to separate carbohydrates using amine compounds as modifier of mobile phase.

In this work, the ultra-performance liquid chromatography (UPLC) with ELSD to quantify the five typical carbohydrates (fructose, glucose, sucrose, lactose, and fructo-oilgosaccharide) was optimized by using the AQUITY UPLC BEH HILIC column and amine compounds as modifiers of mobile phase, and then applied in the analysis of some infant milk powders.

2. Experimental

2.1. Reagents

All reagents were analytical grade. D-Glucose, D-fructose, lactose, sucrose and fructo-oligosaccharide were purchased from Shanghai Biological Chemistry Reagent (China). Acetonitrile (HPLC grade) was purchased from Tedia (USA). Ethanolamine, triethylamine and diethylamine were purchased from Beijing Jingzhan Chemical (China). Ultrapure water was achieved by Millipore Water Purification System (Merck, China). The stock solutions of five typical carbohydrates were prepared by dissolving in aqueous ultrapure water and saved at 4℃. The standard mixture was prepared using these stock solutions. The concentration of each carbohydrate in the standard mixture was 100mg/L, respectively. The standard mixture was used to optimize the separation condition.

## 2.2. Instrument

The ACQUITY[TM] UPLC system (Waters, USA) equipped with ACQUITY UPLC Sample manager (Waters, USA), ACQUITY UPLC Column Heater (Waters, USA) and ACQUITY UPLC ELS detector (Waters, USA) were used to the analysis of carbohydrates. The ACQUITY UPLC® BEH HILIC (2.1mm×150mm, 1.7μm, Waters, Ireland) and VanGuard [TM] Pre-Column(2.1mm×5mm, 1.7μm, Waters, Ireland) were used. The Nitrogen Separator GKG49-1 ( Suzhou gas instrument corporation, China) was used to supply nitrogen gas (99.99%) to ELSD. All the instrument controls, data collection and treatment were performed by MassLynx4.1 (Waters, USA).

## 2.3. Sample preparation

The milk powder for Cow's milk allergy baby (Sample 1) was purchased from Pyongyang Department Store No.1 (DPR of Korea). Other three types of milk powders; YasHILY (China, Sample 2), ArywA (Russia, Sample 3) and Wafari (Singapore, Sample 4) were purchased from the local market. Milk powder samples (0.50g) were weighed exactly, dissolved in 50mL of warm ultrapure water (60~70℃) and maintained for 20min. The solutions were filtered through a 0.45μm membrane filter and then, injected into UPLC system using 10μL loop injector. The sample injection volume was 5μL.

## 2.4. LC Method

The mobile phase for isocratic elution was a mixture of water-acetonitrile (10:90, v/v) containing different concentrations of triethylamine, diethylamine and ethanolamine as modifiers. At First, the mixture of water-acetonitrile (10:90, v/v) containing 0.05% (v/v) tri-ethylamine was used as a mobile phase to separate the carbohydrate standard solution. In the same way, adding ethanolamine, triethylamine and diethylamine to water-acetonitrile (10:90, v/v) solution in different concentrations (0.05~0.1%), the standard mixture was measured. Then, the gradient elution was carried out with mobile phase A (water containing 0.05% triethylamine) and mobile phase B (acetonitrile containing 0.05% ethanolamine and 0.05% triethylamine).

The column temperature was 25℃. A nebulization gas rate was 2.5L/min and a drift tube temperature was 60℃. The gain of ELSD was set to 200. According to the different compositions of mobile phase, the retention time and peak areas of carbohydrates were chosen for the best separation results.

Under this condition, parameters of ELSD were investigated to increase the sensitivity for detection of the carbohydrates. The drift tube temperature was studied in the interval 50~90℃, the nebulization gas rate in 2.0, 2.5, 3.0, 3.5L/min and the photomultiplier tube gain as 50, 100, 200 to compare the peak areas of carbohydrates.

**3. Results and discussion**

**3.1. Optimization of LC separation Condition**

Triethylamine and ethanolamine were already used as modifiers of mobile phase for separation of carbohydrates on silica gel column [20] or BEH amide column [19], and ACQUITY UPLC BEH HILIC column is similar to silica gel column. The amine compound in the eluent can be adsorbed to this HILIC column surface to form a dense monolayer, then carbohydrates can be retained on this dynamically modified amino stationary phase via hydrogen bonding between another amino group of the amine modifiers and hydroxyl group of the carbohydrates [15].

When triethylamine was added into water-acetonitrile mixture (10:90, v/v), the separation between fructose and glucose peaks and between sucrose and lactose peaks were not sufficient in 0.1mL/min of mobile phase flow rate. In the case of diethylamine, the detector response was saturated. The result by using 0.05% (v/v) ethanolamine was the same in triethylamine. Adding both of triethylamine and ethanolamine into water-acetonitrile (10:90, v/v) in 0.05% (v/v) of content respectively, the five carbohydrates were relatively separated, but the separation between fructose and glucose peaks was still not sufficient. When the acetonitrile content was increased to 95% in mobile phase, the fructose and glucose peaks were sufficiently separated, but the lactose and oligosaccharide peaks were not detected within 40min. The increase of triethylamine content in the mobile phase enhanced the separation, but resulted in high detector response noise. Therefore, the gradient eluent conditions were studied with mobile phase A (water containing 0.05% triethylamine) and mobile phase B (acetonitrile containing 0.05% triethylamine and 0.05% ethanolamine). The result showed that the good separation and shorter analysis time was achieved when the content of A was increased from 10% to 20% in 0-15min. In this work, the effect of column temperature on separation of carbohydrates was not studied and the column temperature was set to 25℃, because it was already mentioned on thermal stability of glucose and other sugars by some researchers [21].

**3.2. ELSD parameters**

The main way to improve the sensitivity of ELSD is increasing S/N by reducing the residue of the solvent after evaporation. The residue in drift tube of ELSD depends on the boiling points of the mobile phase and mobile modifiers, and drift tube temperature. Fig. 1 shows the peak areas of the five carbohydrates measured with varying the drift tube temperature from 50 to 90℃.

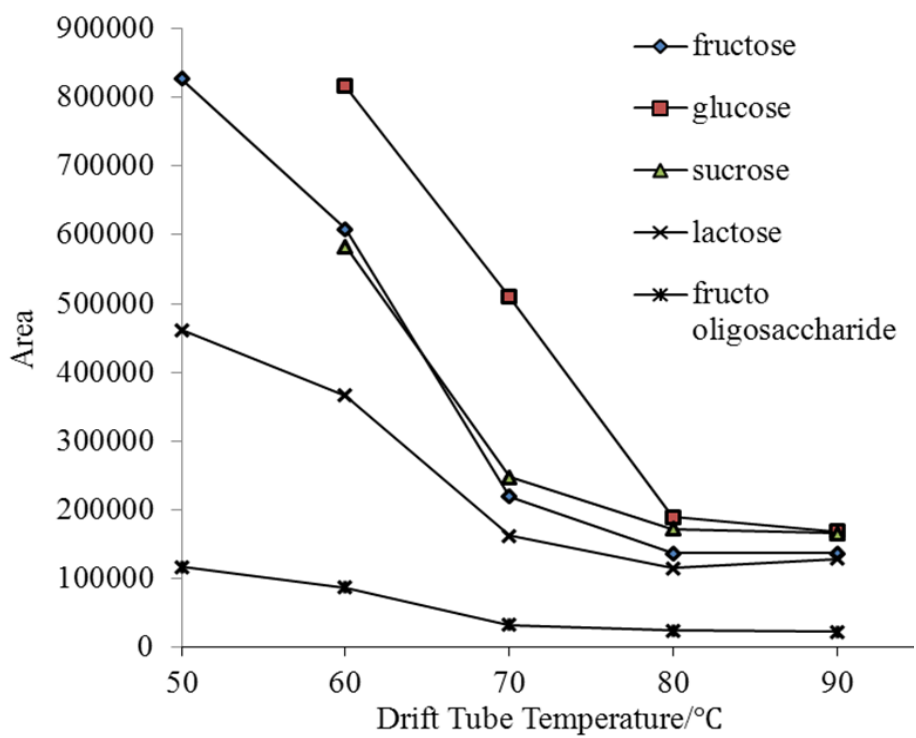

Figure 1.　Peak areas of carbohydrates via drift tube temperature
(column temperature: 25℃, mobile phase A: water containing 0.05% of triethylamine, mobile phase B: acetonitrile containing 0.05% water and 0.05% triethylamine, gradient elution: 0-15min, A:10-20%, nublization gas flow rate: 2.5L/min, gain:200)

As the drift tube temperature increased, all peak areas were decreased and the detector signals were saturated in fructose and glucose at 50℃. Therefore, the adequate temperature was set to 60℃.

Fig. 2 shows the peak areas of carbohydrates and noise levels of detector (noise×1000) measured in 2.0, 2.5, 3.0, 3.5L/min of different nebulization gas flow rate.

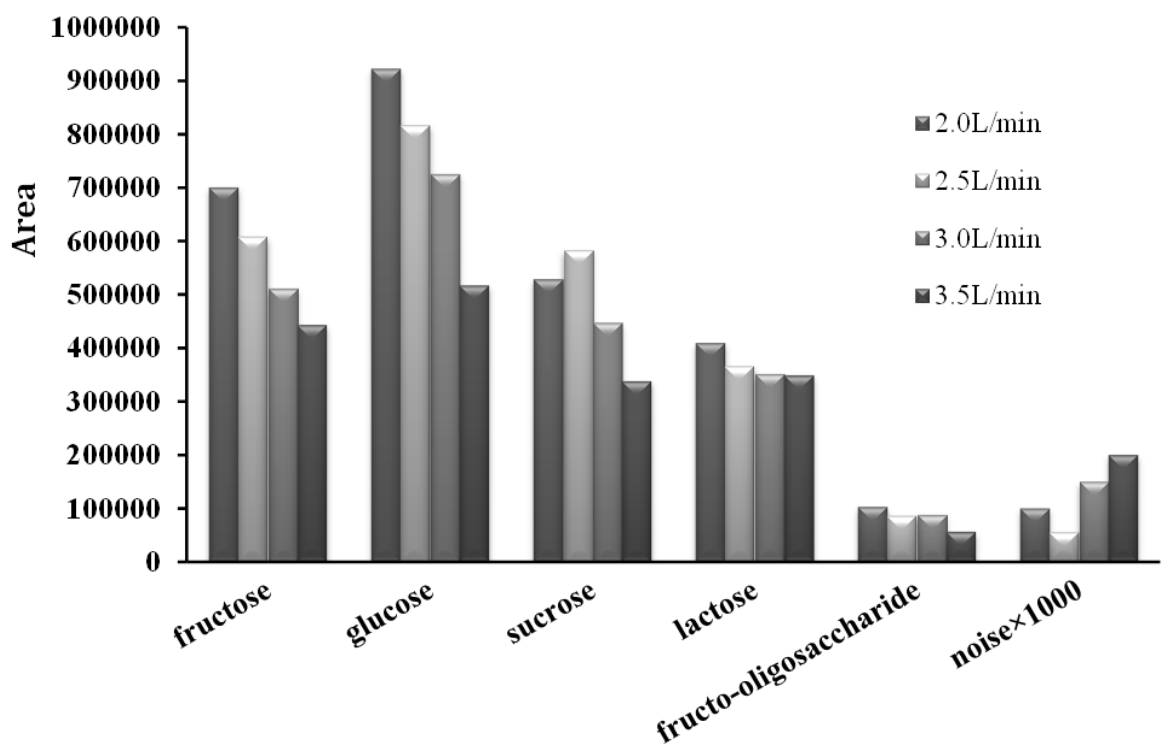

Figure 2. Comparison of peak areas and noise of carbohydrates according to gas flow rate
Column temperature: 25℃, mobile phase A: water containing 0.05% of tri-ethylamine, mobile phase B: acetonitrile containing 0.05% ethanolamine and 0.05% tri-ethylamine, gradient elution: 0-15min, A:10-20%, drift tube temperature 60℃, gain:200

The peak areas were decreased and the noises were increased with the increase of gas flow rate. This shows that the amount of analytes stayed in the drift tube was reduced and the amount of mobile phase was increased in the high gas flow rate. Therefore, the 2.5L/min of gas flow rate was selected.

The photomultiplier tube gain in ELSD is an important factor affecting the limitation of detection (LOD) and the quantification range. If the gain factor is too low, the LOD increases because the signal response decreases, whereas if it is set to higher, the noise also increases, as well as the maximum quantification concentration is lower. ELSD used in this work has the default value of 500 as the best gain. However, since a baseline level becomes higher if the modifier as amine compounds was added to the mobile phase, the 500 of gain value couldn't be used. Therefore, the standard mixture was measured and compared with 50, 100 and 200 of gain. The higher gain resulted in the higher noise level, so the gain of ELSD was set to 100.

Fig. 3 shows the chromatogram of the standard mixture solution measured in 60℃ of drift tube temperature, 2.5L/min of nebulization gas rate and 100 of gain condition.

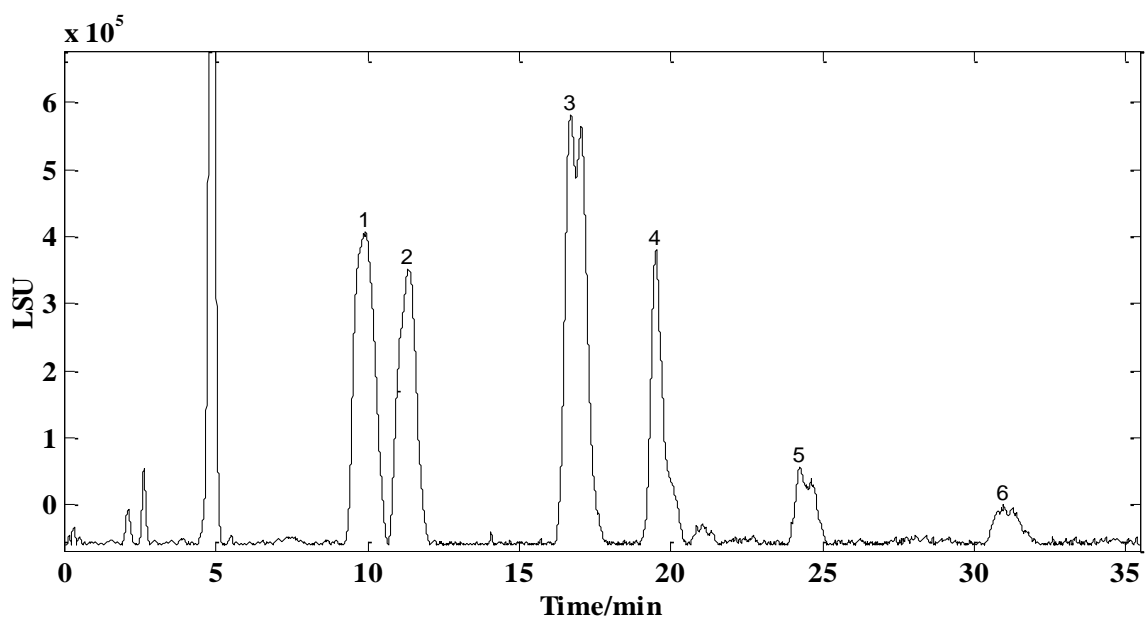

Figure 3. Chromatogram of carbohydrate standard mixture in the optimized separation and ELSD (1-fructose ($t_R$:9.94), 2-glucose ($t_R$:11.34), 3-sucrose ($t_R$:16.66), 4-lactose ($t_R$:19.49), 5,6-fructo-oligosaccharide ($t_R$:24.22, 30.95), column temperature: 25℃, mobile phase A: water containing 0.05% of triethylamine, mobile phase B: acetonitrile containing 0.05% ethanolamine and 0.05% triethylamine, gradient elution: 0-15min, A:10-20%, nublization gas flow rate: 2.5L/min, gain:100, drift tube temperature 60℃）

### 3.3. Quantification of carbohydrates

For carbohydrates quantification, the external standard method was used to calibrate the chromatographic system. The carbohydrate standard solutions with different concentrations (20-300mg/L for fructose and glucose, 20-250mg/L for sucrose and lactose, and 35-180mg/L for fructo-oligosaccharide) were used. Calibration curves between peak areas and the concentration of solution injected into LC system were linear for the five carbohydrates. The results are shown in Table 1.

Table 1. Calibration curves determined for five carbohydrates

| Carbohydrates | Concentration range(mg/L) | Calibration curve | $R^2$ |
|---|---|---|---|
| fructose | 20-300 | y = 4630.8x - 75787 | 0.9945 |
| glucose | 20-300 | y = 4370x - 69376 | 0.9983 |
| sucrose | 20-250 | y = 5598.1x - 99530 | 0.9875 |
| lactose | 20-250 | y = 4354.8x - 112416 | 0.9872 |
| fructo-oligosaccharide | 35-180 | y =2679.1x + 77083 | 0.9966 |

Identification of the carbohydrates in sample solutions was performed by comparison with the retention times of the standards. LOD was estimated as the concentration providing a signal three times higher than the standard deviation of the background noise. The LODs were 16.4mg/L for fructose, 16.4mg/L for glucose, 17.3mg/L for sucrose, 20.0mg/L for lactose, and 46.7mg/L for fructo-oligosaccharide, respectively.

### 3.4. Analysis of samples

Fig. 4 presents the chromatograms of sample solutions prepared from four infant milk powders.

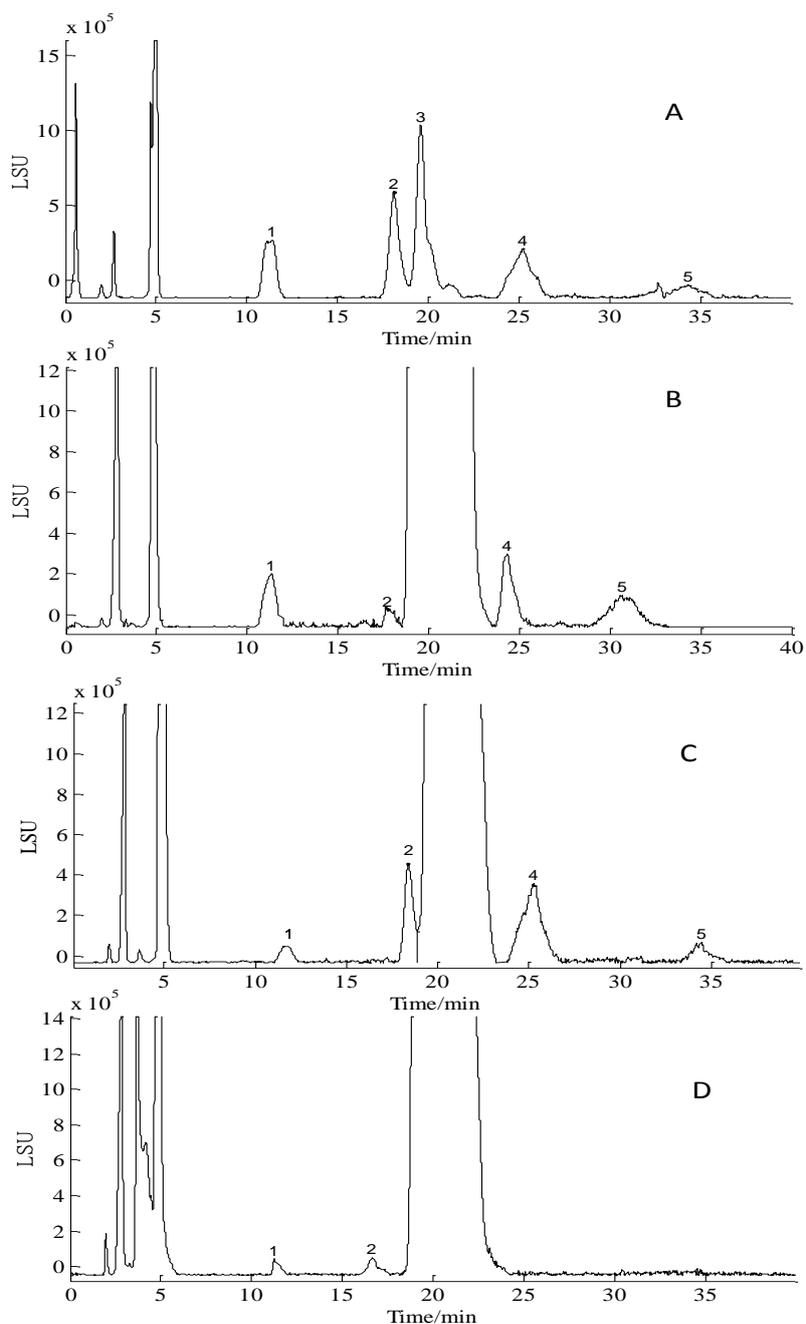

Figure 4.  Chromatograms of four samples measured by ELSD

( A(Sample 1): 1-glucose($t_R$:11.13), 3-lactose($t_R$:19.26), 4- fructo-oligosaccharide($t_R$:25.20),
B(Sample 2): 1-glucose($t_R$:11.33), 3-lactose($t_R$:19.51), 4,5-fructo-oligosaccharide($t_R$:24.33, 30.66),
C(Sample 3): 1-glucose($t_R$:11.73), 3- lactose($t_R$:19.31), 4 -fructo-oligosaccharide($t_R$:25.27),
D(Sample 4): 1-glucose($t_R$:11.24), 2-sucrose($t_R$:16.65), 3-lactose($t_R$:19.50))

All of the milk powder samples had large amount of lactose which gave the saturated response of detector. Therefore, the determination of lactose was carried out by diluting the initial sample 40 times. Fig. 4 A is a chromatogram of the solution diluted 5 times. Sample 1 had relatively low concentration of lactose, because this milk powder was prepared for cow's milk allergy children. In chromatograms of Sample 1-3, there were some peaks with 18.0±0.32 and 34.3±0.10 of retention time. This suggests that these samples have other carbohydrates different from the five ones studied in this work. The contents of carbohydrates in each samples were summarized in Table 2.

Table 2. Determination results of carbohydrates in infant milk power samples (n=5)

| Sample | Glucose/% | Sucrose/% | Lactose/% | Fructo-Oligosaccharide/% |
|---|---|---|---|---|
| 1(A) | 10.37±0.38 | n.d.[*] | 21.83±0.82 | 0.71±0.08 |
| 2(B) | 0.63±0.05 | n.d. | 47.03±3.89 | 1.69±0.15 |
| 3(C) | n.d. | n.d. | 32.96±3.33 | 1.91±0.16 |
| 4(D) | 0.23±0.02 | 0.26±0.06 | 45.87±2.09 | n.d. |

( * n.d. means 'not detected' )

Suitable amounts of the carbohydrates standards ( 2000μg/g sample for fructose and glucose, 1000μg/g sample for lactose, sucrose, and oligosaccharide ) were added to the carbohydrate contents known samples and the proposed analysis procedure was carried out to determine recoveries of each carbohydrate. Recoveries were 95.2-100.2% for fructose, 98.7-103.1% for glucose, 101.0-105.4% for sucrose, 96.4-102.5% for lactose and 95.0-98.4% for fructo-oligosaccharide. The precision of this method was evaluated taking the relative standard deviation (RSD) for 5 analyses of each sample. RSDs were ranged between 3.45-4.23%, 1.46-4.17%, 4.14-5.60%, 1.39-3.53%, and 2.49-3.61% for fructose, glucose, sucrose, lactose, and fructo-oilgosaccharide, respectively.

## 4. Conclusion

The possibility of carbohydrate separation in BEH HILIC column was studied by ultra-performance liquid chromatography with evaporative light scattering detector and mobile phase containing amine compounds as modifiers. Triethylamine and ethanolamine were selected as good modifiers of water-acetonitrile mobile phase, and their concentrations in the mobile phase were optimized to obtain the best separation condition for five carbohydrates such as fructose, glucose, sucrose, lactose, and fructo-oligosaccharide. Four kinds of infant milk powers studied in this work had different carbohydrate compositions and some other carbohydrates different from the five typical ones studied here.

**References**

[1] A. C. Soria, M. L. Sanz, M. Villamiel, Food Chem. 114 (2009) 758.

[2] A. I. Ruiz-Matute, L. Ramos, I. Martinez-Castro, M. L. Sanz, J. Agric. Food Chem. 56 (2008) 8309.

[3] A. Montilla, A. I. Ruiz-Matute, M. L. Sanz, I. Martinez-Castro, M. D. del Castillo, Chromatographia 62 (2005) 311.

[4] L. La Pera, G. Di Bella, R. Magnisi, V. Lo Turco, G. mo Dugo, Ital. J. Food Sci. 19(2007) 319.

[5] E. Dvorackova, M. Snoblova, P. Hrdlicka, J. Sep. Sci. 37 (2014) 323.

[6] L. C. Nogueira, F. Silvab, I. M. P. L. V. O. Ferreirab, L. C. Trugoa, J. Chromatogr. A 1065 (2005) 207.

[7] T. B. Zhang, J. Xu, L. Zhang, W. Zhang, Y. Zhang, J. Pharm. Biomed. Anal. 102 (2015) 1.

[8] M. J. Rojas, T. C. Castral, R. L. C. Giordano, P. W. Tardioli, J. Chromatogr. A 1410(2015) 140.

[9] J. Li, M. Chen, Y. Zhu, J. Chromatogr. A 1155 (2007) 50.

[10] Y. C. Lee, J. Chromatogr. A 720 (1996) 137.

[11] X. Wang, Y. Chen, Carbohydr. Res. 332 (2001) 191.

[12] C. Martinez Montero, M.C. Rodriguez Dodero, D.A. Guillen Sanchez, C. G. Barroso, Chromatographia 59 (2004) 15.

[13] A. Clement, D. Young, C. Brechet, J. Liq. Chromatogr. 15 (1992) 805.

[14] M. Boumahraz, V. Y. Davydov, A.V. Kiselev, Chromatographia 15 (1982) 751.

[15] Y. Wei, M. Y. Ding, J. Chromatogr. A 904(2000) 113.

[16] M. Boumahraz, V. Y. Davydov, A.V. Kiselev, Chromatographia 15 (1982) 751.

[17] E.F. Hounsell, J.M. Rideout, N.J. Pickering, C. K. Lim, J. Liq. Chromatogr. 7 (1984) 661.

[18] C. H. Lochmuller, W. B. Hill, J. Chromatogr. 264 (1983) 215.

[19] N. Dartora, L. M de Souza, A. P Santana-Filho, M. Iacomini, A. T. Valduga, P. A. J. Gorin, G. L. Sassaki, Food Chem. 129 (2011) 1453.

[20] Y. Wei, M. Y. Ding, J. Liq. Chromatogr. 25 (2002) 1769.

[21] R. Slimestad, I. M. Vagen, J. Chromatogr. A 1118 (2006) 281.